\documentclass[12pt,thmsa]{article}
\usepackage{amsmath, latexsym, amsfonts, amssymb, amsthm, amscd}
\usepackage{graphics}
\usepackage{graphicx}

\usepackage{bbold} 

\usepackage[round]{natbib}

\usepackage{lineno}

\begin{document}

\begin{center}
{\Large
\textbf{Sustainable deployment of QTLs conferring\\ quantitative resistance to crops: first lessons\\ 
\vspace*{.05in}
from a stochastic model}}

\vspace*{.3in}

R. Bourget$^{1,2,\ast}$, 
L. Chaumont$^{1}$, 
C.E. Durel$^{2}$,
N. Sapoukhina$^{2}$

\end{center}

\noindent $^1$ LAREMA, D\'epartement de Math\'ematiques, Universit\'e d'Angers, 
2, Blvd Lavoisier Angers Cedex 01, 49045, France\\
$^2$ IRHS (INRA, Universit\'e d'Angers, Agrocampus Ouest), SFR QUASAV, 
rue G. Morel F-49071 Beaucouz\'e, France\\
$\ast$ E-mail: bourget.romain@gmail.com

\section*{Summary} 
\begin{itemize}

\item[$\bullet$] Quantitative plant disease resistance is believed to be more durable than qualitative resistance, since it exerts less selective pressure on the pathogens. However, the process of progressive pathogen adaptation to quantitative resistance is poorly understood, which makes it difficult to predict its durability or to derive principles for its sustainable deployment. Here, we study the dynamics of pathogen adaptation in response to quantitative plant resistance affecting pathogen reproduction rate and its carrying capacity.

\item[$\bullet$] We developed a stochastic model for the continuous evolution of a pathogen population within a quantitatively resistant host. We assumed that pathogen can adapt to a host by the progressive restoration of reproduction rate or of carrying capacity, or of both. 
\item[$\bullet$] Our model suggests that a combination of QTLs affecting distinct pathogen traits was more durable if the evolution of repressed traits was antagonistic. Otherwise, quantitative resistance that depressed only pathogen reproduction was more durable. 

\item[$\bullet$] In order to decelerate the progressive pathogen adaptation, QTLs that decrease the pathogen's ability to extend must be combined with QTLs that decrease the spore production per lesion or the infection efficiency or that increase the latent period. Our theoretical framework can help breeders to develop principles for sustainable deployment of quantitative trait loci.
\end{itemize}

\vspace*{.1in}

\noindent {\bf Key words: } Continuous adaptive dynamics, birth and death process, QTL erosion, resistance durability, partial resistance, breeding strategies.
\vspace{1cm}
\section{Introduction}

The use of resistant disease cultivars in agro-ecosystems has a considerable effect on the evolutionary dynamics of pathogens \citep{Mcd2002,Mun2002}. Two types of resistance have historically been distinguished: qualitative and quantitative resistance \citep{Ain1981}. Qualitative resistance is based on gene-for-gene relationships subjected to relatively simple genetic control, and it renders a cultivar immune to disease \citep{Flo1971,Dan2001}. The mechanisms of the rapid adaptation of the pathogen to this type of resistance, and the evolutionary consequences of its deployment, have already been studied in detail \citep{Bos2003, Bou2013}. In contrast, the genetic mechanisms underlying pathogen adaptation to quantitative plant resistance, which reduces the level of disease rather than conferring immunity \citep{You1996}, are poorly understood. It has been shown that quantitative resistance in a host population can be associated with the presence of quantitative trait loci, QTLs, in its genome \citep{You1996,Cla2010}. QTLs can be overcome by a swift pathogen adaptation like a qualitative resistance \citep{Leh1996,Mou2004,Pal2009}. In this case pathogen adaptation can be described by existing models \citep{Bou2013}. Here, we consider a progressive pathogen adaptation when the polygenic structure is overcome by multiple successive genetic changes in the pathogen population \citep{Mcd2002,Pol2009} that can make quantitative resistance more durable \citep{And2007,Pal2009,Bru2010}. Moreover, it has been shown that quantitative resistance can depress distinct life-history traits of the pathogen population, such as the latent period, infection efficiency, lesion size or the spore production rate \citep{Par2009,Azz2012,Lan2012,Ber2013}. However, no study, whether theoretical or empirical, has demonstrated how the ways of restoring the various pathogen traits could drive the speed of pathogen adaptation. The lack of empirical studies can be explained by the fact that observation of the progressive adaptation by the pathogen to quantitative resistance is a labor- and time-consuming process. Indeed, the corresponding pathogen populations can display a distribution of pathogenicity that may vary considerably from year to year as a result of the strong genotype-by-environment interactions that occur for quantitative characters in both the plant and the pathogen \citep{Mcd2002,Caf2013}. Thus, to be able to predict the durability of quantitative resistance, we need a theoretical approach to explore the process of progressive pathogen adaptation driven by the stochastic restoration of depressed life-traits. \\

In recent years, growing attention has been paid to approaches in which epidemiology and evolutionary ecology are merged to study the evolution of plant pathogens in response to various different deployments of quantitative plant resistance. For instance, \cite{Leh1996} have demonstrated that quantitative resistance will select pathogens with a shorter latent period. Using a deterministic model, \citet{Iac2012} have shown that the increase in the proportion of hosts carrying quantitative resistance increases the gain in the duration of the healthy canopy area. They have also shown that quantitative resistance that reduces the infection efficiency gives a greater gain than quantitative resistance that reduces the sporulation rate. Theoretical approaches investigating the evolutionary consequences of partially effective (imperfect) vaccines have focused on two main questions: how an imperfect vaccine selects for pathogens that are able to evade the protective effects of the vaccine \citep{Bov2005}, and how an imperfect vaccine causes evolutionary changes in pathogen virulence defined as host mortality caused by the pathogen \citep{Gan2001}. Recently, \citet{Gan2007} have developed a model by which to consider a set of pathogen strains that arise and spread in the face of vaccination and to classify them by their life-history parameters, rather than to distinguish only escape and virulent mutants within the pathogen population. However, none of these models has provided a realistic description of continuous pathogen adaptation within a host carrying quantitative resistance. \\

The objective of this paper is to study the dynamics of continuous pathogen adaptation to a quantitatively resistant host. To do this, we adapted and analyzed a stochastic model \citep{Cha2006} that describes the birth, mutation, and death processes of pathogen lesions on a quantitatively resistant host, underlying the dynamics of the pathogen population. The model made it possible to classify pathogen strains by their adaptation state and to track the adaptation dynamics of strains through the progressive stochastic changes in evolution coefficients mimicking the restoration of the pathogens life-history parameters that have been depressed by quantitative resistance. By numerical simulations, we studied pathogen adaptation dynamics in response to different impacts of quantitative resistance on distinct pathogen life-history parameters. Our model suggested that a combination of QTLs affecting distinct pathogen traits was durable, if the restoration process of repressed traits was antagonistic or independent. Otherwise, quantitative resistance that depressed pathogen reproduction alone was more durable. We discuss strategies for the judicious use of quantitative trait loci in plant breeding from the viewpoint of potential pathogen adaptation.

\section{Description}
\subsection{Model overview}
Our model describes the evolutionary dynamics of a polycyclic disease on an individual host. We consider the pathogen population as a set of lesions that can be produced by genetically different strains. We assume that each pathogen lesion interacting with an individual host is characterized by two quantitative traits of pathogenicity: the reproduction rate, that is the number of lesions produced by a lesion per unit of time, and the pathogen's carrying capacity, that is the maximum number of lesions that can coexist on a given host \citep{Ott2007}. The reproduction rate combines the spore production per lesion, the infection efficiency and the latent period. The magnitudes of both the reproduction rate of the pathogen and its carrying capacity depend on the pathogen characteristics, the level of host resistance as well as host-pathogen interactions \citep{Lan2012}. Despite this fact, for the sake of simplicity, hereinafter we refer to pathogen reproduction rate and its carrying capacity as pathogen traits, since we assume that only the pathogen population evolves and the host surface exposed to infection is fixed.

We assume that on a susceptible host, the reproduction rate of each lesion is $r$, and the maximum number of lesions that can potentially coexist is $K$. If the pathogen attacks a quantitatively resistant host, the host resistance depresses pathogen traits to some extent. Thus, on a quantitatively resistant host, the pathogen reproduction rate is reduced to $\beta_r r \leq r$ and its carrying capacity is reduced to $\beta_K K \leq K$, where $\beta_r, \beta_K \in ]0,1]$. We assume that the pathogen is able to adapt to the host resistance by a progressive restoration of the depressed traits to the original values, $r$ and $K$, corresponding to a susceptible host. When depressed pathogen traits attain the original values, the quantitative host resistance is supposed to be completely eroded or overcome. In our model, the evolution of pathogen traits on a quantitatively resistant host is assumed to be driven by stochastic mutations and selection. We assume that the lesions have the same size and that this size is constant. Thus, an increase in the carrying capacity, $\beta_K K$, leads to the extension of the maximum infected host surface. We designate $\beta_r$ and $\beta_K$ as evolution coefficients of pathogen traits, since whether they are raised or lowered by stochastic mutations, they drive the evolution of the reproduction rate and carrying capacity, respectively. Thus, the values of evolution coefficients describe the adaptation state of a pathogen lesion. The higher the values of $\beta_r$ and $\beta_K$, the closer the abilities of the pathogen to those on a susceptible host and consequently the pathogen strain becomes more adapted to a partially resistant host. The emergence of new pathogen strains with different pathogen traits, $\beta_r r$ and $\beta_K K$, leads to a diversification of the pathogen population and, thereby, to inter-strain competition and selection.

\subsection{Individual-based stochastic model for progressive pathogen adaptation}
Our model is an adjustment of the theoretical framework of \citet{Cha2006} based on Markov chains, which captures stochastic adaptive dynamics influenced by a continuous trait, as in the case of continuous pathogen adaptation to quantitative host resistance. Being individual-based, our model makes it possible to follow the adaptation state of each lesion and the stochastic dynamics of the number of lesions of each pathogen strain infecting an individual host carrying quantitative resistance. Here, the strain is a set of lesions having the same adaptation state. Let $I_n(t)$ be a random variable for the number of pathogen lesions making up the $n$th strain at time $t$. Thus, $I(t) = \sum \limits_{n=1}^{N(t)} I_n(t)$ is a random variable for population size, that is the total number of pathogen lesions on the host at time $t$, where $N(t)$ is the total number of strains in the pathogen population. Each lesion, $i \in [1,I(t)]$, is characterized by a vector of adaptation state $\vec{\beta^i}=(\beta_r^i,\beta_K^i)$ $\in ]0;1]^2$, where $\beta_r^i$ and $\beta_K^i$ are the evolution coefficients of traits $r$ and $K$, respectively. The dynamics of the number of lesions of the $n$th strain, $I_n(t)$, are driven by three major processes: birth, death and mutation. These stochastic processes increase or decrease the variable by one. The transition from one state of the variable $I_n(t)$ to another $I_n(t) \pm 1$ at the moment of time $t$ is driven by transition functions \citep{Nai1993,All2003}. For each lesion $i$ with evolution vector $\vec{\beta^i}$, we define the birth transition function without mutation, $\lambda(\vec{\beta^i})$, the death transition function due to competition, $\mu(\vec{\beta^i})$, and the birth transition function with mutation 

$\gamma(\vec{\beta^i})$ :
\begin{equation}\label{jump}
\left\{ \begin{array}{llll}
\lambda(\vec{\beta^i})&=&\beta_r^i r(1-\omega) \\
\mu(\vec{\beta^i})&=&\frac{\beta_r^i r}{\beta_K^i K}\left(\frac{\sum \limits_{j=1}^{I(t)} (\beta_r^j+\beta_K^j)}{\beta_r^i+\beta_K^i}-1\right)\\
\gamma(\vec{\beta^i})&=& \beta_r^i r \omega \\
\end{array}\right.,
\end{equation}

where $\omega$ is the probability for a lesion to create a mutant lesion generating a new strain. The product of the pathogen reproduction rate, $\beta^i_r r$, and the probability of not mutating, $(1-\omega)$, gives us the birth transition function of lesion $i$, $\lambda(\vec{\beta^i})$, while with the probability of mutating, $\omega$, it gives us its mutation function, $\gamma(\vec{\beta^i})$. The death function is based on the Lotka-Volterra equation \citep{Ott2007}, where the competition between two lesions, $i$ and $j$, is determined by the ratio between their mean evolution coefficients $ (\beta_r^j+\beta_K^j) / ( \beta_r^i+\beta_K^i)$, while the carrying capacity of lesion $i$ is $\beta^i_K K$. The term $\beta^i_r r$ of the death rate leads to equal birth and death rates, when all lesions have the same evolution coefficients and the number of lesions reaches its maximum, $\beta^i_K K$. Table \ref{param} summarizes the variables and parameters of the model. 

Despite the fact that our model is continuous in time, for numerical simulations we used time discretization with random step, $t_x$. We assume that at time $t+t_x$, only one of the events birth, death or mutation, can occur. To determine the time step $t_x$ and what event happens to what pathogen lesion, we generate for each lesion $i\in [1,I(t)]$ three independent random numbers from the exponential distributions with parameters $\lambda(\vec{\beta^i})$, $\mu(\vec{\beta^i})$, $\gamma(\vec{\beta^i})$, corresponding to birth, death and mutation events, respectively. Then, $t_x$ becomes the smallest random number among $3I(t)$ generated ones. Note that $t_x$ follows an exponential distribution with parameter $\sum \limits_{i=0}^{I(t)} \lambda(\vec{\beta^i})+\mu(\vec{\beta^i}) + \gamma(\vec{\beta^i})$. We also memorize the event and number of lesion $i$ corresponding to $t_x$. Thus, the time advances to $t+t_x$ and the corresponding event happens for pathogen lesion $i$. 

\begin{figure}[h]
\begin{center}
\includegraphics[width=16cm]{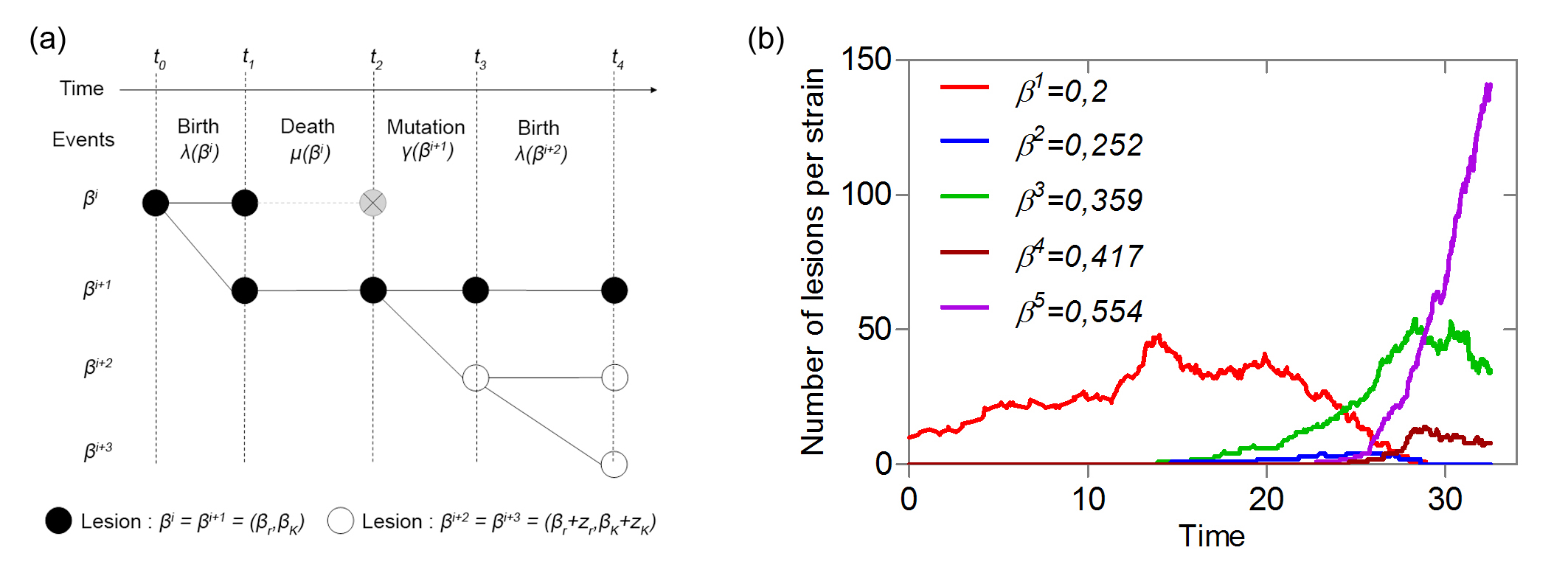}
\end{center}
\caption{
{\it {\small {\bf Model behavior.} (a) Stochastic events with their transition functions (2.1) driving continuous adaptation of the pathogen population to a quantitatively resistant host. We follow stochastic dynamics of the number of pathogen lesions per strain. Let us assume that the adaptation process starts at time $t=t_0$ with one lesion with an evolution vector $\beta^i = (\beta_r,\beta_K)$. Mutation at $t=t_3$ leads to an emergence of a new strain with an evolution vector $\beta^{i+2} = (\beta_r+z_r,\beta_K+z_K)$ with $(z_r,z_K)$ picked randomly according to a scenario of the pathogen adaptation (part \ref{scenar}). The interval between two consecutive times $t_y$ and $t_{y+1}$ is the time before an event happens. (b) Sample path of stochastic model (2.1). During the adaptation driven by identical (r,K)-restoration, the pathogen population splits into five distinct strains characterized by evolution coefficients $\beta^1, \dots, \beta^5$, which population sizes are driven by stochastic events. Due to inter-strain competition, the pathogen strain with the highest evolution coefficient, $\beta^5$, is selected. The parameter values and initial conditions are $I_1(0)=10$, $r=1$, $b=0.2$, $\omega = 0.05$, $K=1000$.}}}
\label{Fig1}
\end{figure}

If a birth event happens for lesion $i$ with evolution vector $\vec{\beta^i}$, then this lesion produces a new lesion with the same evolution coefficients, $(\beta^i_r, \beta^i_K)$. If a death event happens for lesion $i$, this lesion dies because of the competition with the other lesions. If a mutation event happens for lesion $i$, a new lesion corresponding to a genetically distinct strain appears with a new evolution coefficient $\vec{\beta^i}+\vec{z} =(\beta^i_r+z_r, \beta^i_K+z_k)$, where random $\vec{z} =(z_r,z_k)$ is determined according to scenarios of pathogen adaptation, described below in part \ref{scenar}. Then, we again generate $3I(t)$ random numbers and repeat the iterations listed above. The cycle is repeated until the convergence of one of the evolution coefficients, $\beta^i_r$ or $\beta^i_K$ , to 1. Figure \ref{Fig1} summarizes the behavior of the model in terms of the transitions possible in a small time interval, $t_x$. 

\subsection{Scenarios of pathogen adaptation}\label{scenar}
Let us consider five different scenarios of pathogen adaptation. Recall that the adaptation process is driven by random evolution coefficients. The distribution law of evolution coefficients seems to be relatively tightly distributed around the initial coefficients \citep{Mar2006,San2010b}. Thus, for all the adaptation scenarios, we considered a uniform mutation law that truncates $\vec{\beta^i}+\vec{z}=(\beta^i_r+z_r,\beta^i_K+z_K)$ in order to keep the evolution coefficients within $]0,1]^2$. This means that after a mutant birth, when picking mutant coefficients $\beta^i_r +z_r$ or $\beta^i_K + z_K > 1$ (or $\beta^i_r +z_r$ or $\beta^i_K + z_K < 0$), $\beta^i_r +z_r$ or $\beta^i_K + z_K$ is set to $1$ (or to 0.0001, respectively).

Let us formally define the five possible adaptation scenarios for pathogen lesion $i$ (for a more detailed description of mutation laws see Supporting Information Notes S1):
\begin{itemize}
\item[($1$)] Identical $(r, K)$-restoration: when both depressed traits, i.e. $\beta_r^ir<r$ and $\beta_K^iK < K$, evolve with the same evolution coefficients, that is for all lesions $i$ at any time $t$, $\beta_r^i=\beta_K^i=\beta^i$. A pathogen lesion evolving through a reproduction creates a new lesion $j$ of a new pathogen strain with the evolution coefficients $\beta_r^j=\beta_K^j=\beta^i+z$. Here, $z$ follows a uniform law in $[-b,b]$ where $b$ is the maximum rise and fall of the evolution coefficient, induced by a mutation.

\item[($2-3$)] $r$- (or $K$-)restoration: when only one trait is depressed by host resistance, $\beta_r^ir<r$ (or $\beta_K^iK < K$), and evolves during adaptation. Thus, for all lesions $i$, $\beta_r^i$ (or $\beta_K^i$) can evolve, and $\beta_K^i$ (or $\beta_r^i$) stays constant at $1$. Then, a pathogen lesion evolving through a reproduction creates a new lesion $j$ of a new pathogen strain with the evolution coefficients $\beta_r^j=\beta_r^i+z$ (or $\beta_K^j=\beta_K^i+z$) and $\beta_K^i$ (or $\beta_r^i$) stays constant at $1$. Here, $z$ follows a uniform law in $[-b,b]$.
\item[($4$)] Independent $(r, K)$-restoration: when both depressed traits evolve independently. Then, a pathogen lesion evolving through a reproduction creates a new lesion $j$ of a new pathogen strain with the evolution coefficients $\beta_r^j=\beta_r^i+z_r$ and $\beta_K^j=\beta_K^i$ or $\beta_r^j=\beta_r^i$ and $\beta_K^j=\beta_K^i+z_K$, $z_r$ and $z_K$ are two independent random variables following a uniform law in $[-b,b]$.

\item[($5$)] Antagonistic $(r, K)$-restoration: when the evolution of both depressed traits is determined by a trade-off between them. Then, a pathogen lesion evolving through a reproduction creates a new lesion $j$ of a new pathogen strain with the evolution coefficients $\beta_r^j=\beta_r^i+z_r$ and $\beta_K^j=\beta_K^i + z_K$. With a probability $1/2$ we picked $z_r$ uniformly in $[- (8/5)b,(8/5)b]$, and we fixed $z_K= - z_r/2$, and with a probability $1/2$ we picked $z_K$ uniformly in $[-(8/5)b,(8/5)b]$ and we fixed $z_r= - z_K/2$. Intervals were set in order to keep the mutation variances equal to those from the other scenarios.
\end{itemize}

To test the sensitivity of the model to mutation laws, we considered two different stochastic laws for the case of the adaptation driven by identical $(r, K)$-restoration, that is $\beta_r^i=\beta_K^i=\beta^i$. Firstly, we assume $z$ to follow a truncated normal law, $\mathcal{N}(0,b/\sqrt{3})$. As for the antagonistic $(r, K)$-restoration process the variance is set to be equal to that from the uniform stochastic law. Secondly, $z$ follows a truncated uniform scaling law, $\mathcal{U}_{[-2b(1.1-\beta^i),2b(1.1-\beta^i)]}$, representing a situation when at the beginning of the process pathogen lesions are poorly adapted and so most mutations are advantageous, but then, when the individuals have almost completely adapted, advantageous mutations become uncommon, as observed in \citet{Nov1995}. Since the variance changes during the simulation, it is the only situation where the variance is not equal to that from the truncated uniform law.

\subsection{Model implementation}
For all adaptation scenarios, at the beginning of the adaptation process, the pathogen population consisted of 10 lesions of a single strain, $I_1(0)=10$. We assumed that these lesions interacted with a host carrying quantitative resistance. Thus, for each of the adaptation scenarios, we defined the initial values of evolution coefficients as equal to $0.2$ for depressed traits and $1$ for non-depressed traits. The established initial conditions correspond to a primary infection of a quantitatively-resistant host by a single pathogen strain. 

To study the speed of pathogen adaptation, 1000 simulations were run for every parameter set. A simulation stopped when its time reached $1.1T$, where $T$ was the first time when the population evolution coefficient of one of the depressed traits reached $0.9$.

For each run $k \in {1,2,...,1000}$, we determined dynamics of the total population size,
$I^k(t_y^k)$ and the dynamics of the population evolution vector over the whole population,

\[\bar{\beta_r^k}(t_y^k)=\sum_{i=1}^{I(t_y^k)} \beta_r^i/I(t_y^k), \mbox{
and } \bar{\beta_K^k}(t_y^k)=\sum_{i=1}^{I(t_y^k)} \beta_K^i/I(t_y^k),\]
where $t_y^k$ is the time at the $y$th event of the run $k$.

Then, we calculated the mean population size over 1000 runs,
\[\bar{I}(\bar{t_y})=\sum_{k=1}^{1000} I^k(t_y^k)/1000,\]

and the mean evolution vector,

\[\bar{\bar{\beta_r}}(\bar{t_y})=\sum_{k=1}^{1000}
\bar{\beta_r^k}(t_y^k)/1000, \mbox{ and }
\bar{\bar{\beta_K}}(\bar{t_y})=\sum_{k=1}^{1000}
\bar{\beta_K^k}(t_y^k)/1000,\]

where $\bar{t_y}= \sum_{k=1}^{1000} t_y^k/1000$ is the mean time of the $y$th event over 1000 runs. 

We applied the Gillespie algorithm \citep{Gil1977} to track the exact trajectories of the Markov chain, and the model was implemented in C++ using Code Blocks and a GNU GCC compiler. To perform numerical simulations, we used a range of parameters sweeping a large spectrum of biological situations. We studied the speed of the adaptation process in the five adaptation scenarios with the following parameter values: $r=1$, $K=1000$ and $b=0.2$. To check the robustness of the results obtained, we performed numerical simulations where we varied the mutation law and values of the $r$, $K$ and $b$ parameters. With identical $(r, K)$-restoration with a truncated uniform law as the mutation law, $r$ were varied in $\{0.5, 1, 2, 3\}$, for each fixed carrying capacity, $K$ $\in\{100,500,1000,2000\}$ and with $b$ taking values in $\{0.1,0.2,0.3\}$. We also examined model behavior for two pairs of $r$ and $K$: $r=0.5$ and $K=2000$, and $r=2$ and $K=500$ (not illustrated). We tested the normal truncated law and a uniform scaling law with $r=1$, $K=1000$ and $b=0.2$. In all simulations $\omega$ was fixed to $0.05$.

\section{Results}
\subsection{Progressive pathogen adaptation by the restoration of its life-traits}
\begin{figure}[h]
\begin{center}
\includegraphics[height=10cm]{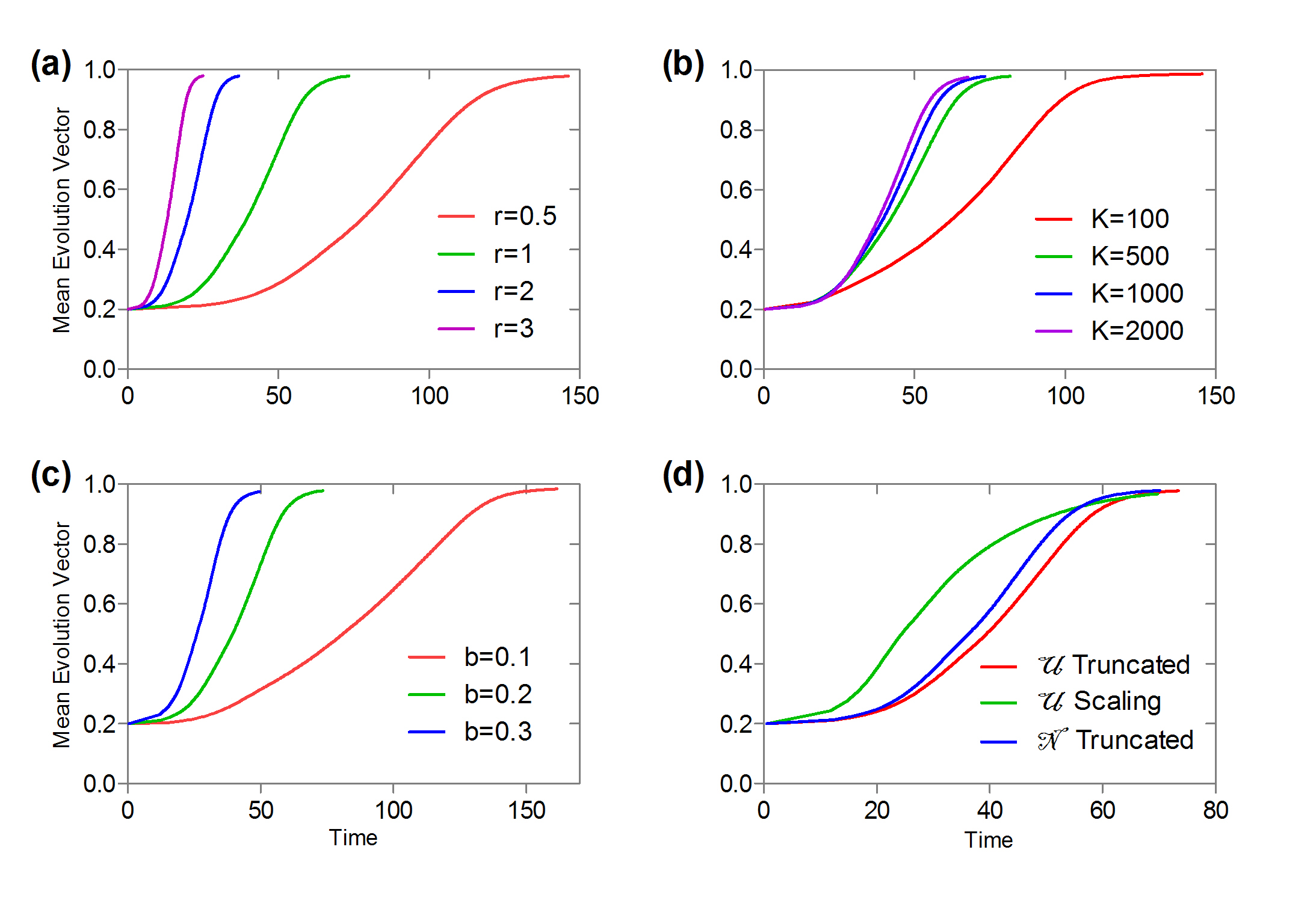}
\end{center}
\caption{
{\it {\small {\bf Evolution of pathogen reproduction rate and its carrying capacity in five scenarios of pathogen adaptation to quantitative resistance.} A continuous adaptation process is described in terms of the dynamics of mean evolution coefficients, $\bar{\bar{\beta_r}}$ , $\bar{\bar{\beta_K}}$ (a) and of the mean size of the pathogen population (b). In (a), unbroken curves represent the dynamics of the mean evolution coefficient of the carrying capacity, $\bar{\bar{\beta_K}}$, dashed curves - the dynamics of the mean evolution coefficient of the reproduction, $\bar{\bar{\beta_r}}$, and dotted curves - the dynamics of both evolution coefficients when they are equal, $\bar{\bar{\beta}}=\bar{\bar{\beta_r}}=\bar{\bar{\beta_K}}$. The different colors represent five adaptation scenarios. The inset figure shows the first 30 time units. Other parameters: $I_1(0)=10$, $r=1$, $b=0.2$, $\omega = 0.05$, $K=1000$. }}}
\label{Fig32}
\end{figure}

To study the speed of pathogen adaptation to a partially resistant host population, we considered five scenarios: identical $(r, K)$-restoration process, $r$- or $K$-restoration, independent $(r, K)$-restoration and antagonistic $(r, K)$-restoration. Numerical simulations showed that the dynamics of the mean evolution coefficients and the growth of the mean size of the pathogen population were increasing sigmoid functions for every adaptation scenario (Fig. \ref{Fig32}). If the pathogen adapts to a quantitative resistance by the evolution of its carrying capacity, $K$, the adaptation will proceed at the highest speed (Fig. \ref{Fig32}). Moreover, the establishment stage is very short, and the dynamics of the mean evolution coefficient, $\bar{\bar{\beta_K}}$, soon become a quasi-linear function. In this situation, the growth of the mean number of lesions also occurs quickly, but the mean number of lesions is higher at the end of the adaptation process in the $r$-restoration case (Fig. \ref{Fig32}b). The lowest adaptation speed and growth of mean number of lesions correspond to adaptation by the antagonistic restoration of $r$ and $K$ traits. The curves of the dynamics of the mean evolution coefficients corresponding to the cases of identical $(r, K)$-restoration, and $r$-restoration, and the independent restoration of both traits are located between the extreme curves in decreasing order of the adaptation speed. Surprisingly, the speed of the adaptation by identical $(r, K)$-restoration exceeds that by $r$-restoration. Apart from $r$-restoration, the growth of the number of lesions is of the same order, since it corresponds to the growth of the adaptation curves of the carrying capacity (Fig. \ref{Fig32}b). Note that $(r, K)$-restoration with independent or antagonistic evolution coefficients is driven by the restoration of the $r$-trait at the beginning of the process (Fig. \ref{Fig32}a), and subsequently by the restoration of the $K$-trait. In the case of antagonistic evolution coefficients, the restoration process of the reproduction rate is much slower. 

\subsection{The impact of model parameters on adaptation by the identical $(r, K)$-restoration}
\begin{figure}[h]
\begin{center}
\includegraphics[height=10cm]{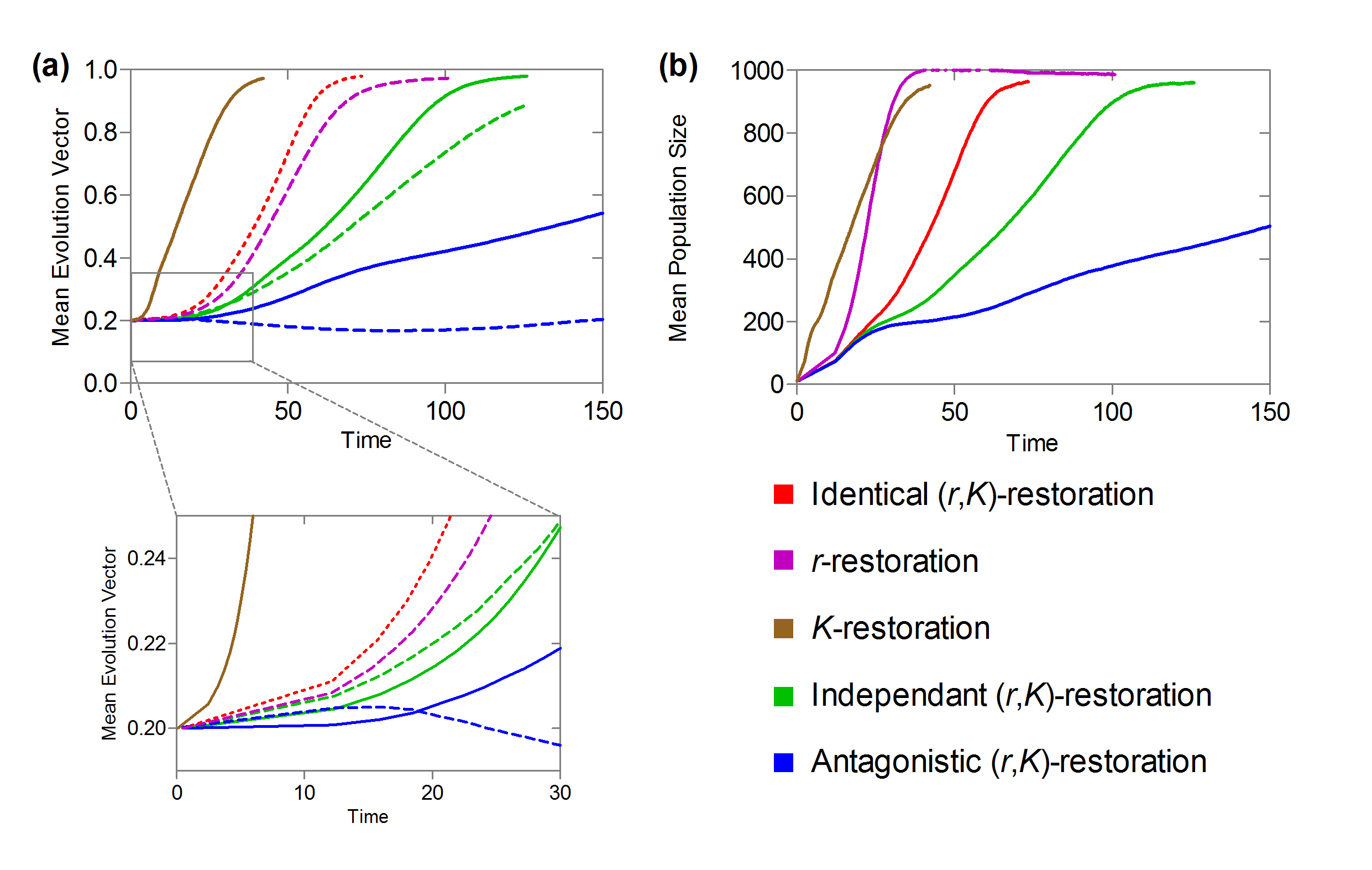}
\end{center}
\caption{
{\it {\small {\bf The impact of the model parameters on the dynamics of adaptation driven by identical ($r,K$)-restoration.} (a) Impact of variation of the reproduction rate, $r$, on the adaptation dynamics. Parameter values: $K=1000$, $b=0.2$. The mutation law is a truncated uniform law; (b) Impact of variation of the carrying capacity, $K$, on the adaption dynamics. Parameter values: $r=1$, $b=0.2$. The mutation law is a truncated uniform law; (c) Impact of variation of the uniform law boundaries, $b$, on the adaptation dynamics. Parameter values: $r=1$, $K=1000$. The mutation law is a truncated uniform law; (d) Impact of variation of the mutation law on the adaption dynamics. Parameter values: $r=1$, $\omega = 0.05$, $K=1000$, $b=0.2$. Other parameters for (a-d): $\beta_r= \beta_K$, $I_1(0)=10$, $\omega = 0.05.$}}}
\label{Fig12}
\end{figure}

First, we studied the impact of three parameters, $r$, $K$, and $b$ on the adaptation process that resulted from identical $(r, K)$-restoration. Numerical simulations showed that the dynamics of the mean evolution coefficients retained a sigmoid shape for any parameter values (Fig. \ref{Fig12}a-c). The higher the values of $r$, $K$, or $b$, the faster the growth of the mean evolution coefficients, which implies rapid pathogen adaptation. Since $r$ multiplies all transition functions, its impact is obvious: variation of $r$ changes the time scale only. In contrast to $r$ and $b$, the increase in the carrying capacity, $K$, has a saturating effect on the speed of pathogen adaptation (Fig. \ref{Fig12}b). $K$ values above a certain threshold have a negligible impact on the speed of adaptation. \\

Analysis of the effects of the mutation laws on the adaptation process demonstrated that for each of the laws considered, the mean evolution coefficients grew continuously, changing in shape from the sigmoid curve obtained for normal and truncated uniform laws to the concave shape obtained for a uniform scaling law (Fig. \ref{Fig12}d). Compared to the truncated uniform law, a normal distribution of mutation slightly accelerates the adaptation process.

\section{Discussion}
In this article, we describe a stochastic framework developed to study the speed of pathogen adaptation to quantitative plant resistance. The model allows us to provide some general guidance about how to manage quantitative resistance so as to increase its durability. In particular, we show that in order to decelerate the progressive pathogen adaptation, QTLs that decrease the pathogen's ability to extend must be combined with QTLs that decrease its rate of reproduction.\\

\subsection{Sustainable use of QTLs that confer quantitative resistance to cultivars}
The shape of the adaptation curve is crucial for predicting the durability of the quantitative resistance, since it allows us to predict the speed of the adaptation process. \citet{Nov1995} showed that the fitness of RNA viruses increases exponentially. Nevertheless, it has been shown that the experimental method used did not reflect the epidemiological aspects of parasite evolution, since it did not consider the transmission of disease from host to host \citep{Ebe1998}. Our model predicts that progressive pathogen adaptation is an S-shaped function with establishment, growth, and saturation stages. The growth stage corresponds to the growth of the effective population size. Indeed, as the number of lesions increases, the number of emerging pathogen mutants with a restored trait increases, which speeds up the adaptation process, as pointed out by \citet{Ebe1998}. The saturation stage, which corresponds to the deceleration of the adaptation process, occurs when the adaptation is no longer driven by beneficial mutations, but rather by the selection of strains with the highest evolution coefficients. Moreover, the higher the mean evolution coefficient, the harder it is for a beneficial mutant to be fixed within the pathogen population. In our within-host model, disease transmission among hosts is not considered either, but the limitation of the evolution coefficient to a value of less than one can explain the saturation that occurs following exponential growth and that results in an S-shaped function. To validate the shape of the theoretical adaptation curve we need experimental data for the long-term dynamics of pathogen fitness, or for the evolution of quantitative traits of pathogenicity.\\

Moreover, we showed that the pathogen adaptation is slightly slower when partially resistant cultivars control only the pathogen reproduction than when they reduce both the pathogen reproduction rate and its carrying capacity, if the adaptation of both traits proceeds at the same speed. This finding can be explained by the level of selection in a pathogen population. When the partial resistance reduces only the pathogen reproduction rate, strains can easily extend. Conversely, when partial resistance of the host reduces both the pathogen reproduction rate and its carrying capacity, pathogen strains compete with one another within a host, leading to strong selection for the better adapted strain and acceleration of the adaptation process. Nevertheless, if the resistance affects only the pathogen reproduction, the pathogen is still easily able to extend. This soon renders the resistance ineffective at limiting pathogen invasion.\\

Recent empirical studies have demonstrated the existence of independent and antagonistic evolution of depressed pathogen traits in an adaptation process \citep{Azz2012,Azz2013}. Our model suggests that the most durable combinations of QTLs are those that induce an antagonistic or an independent restoration of repressed reproduction rate and pathogen carrying capacity. This finding confirms the hypotheses of \citet{Par2009}, suggesting that trade-offs between traits of pathogenicity might constrain the pathogen adaptation to quantitative trait loci, since they apply divergent selective pressures to the pathogen population. Thus, our theoretical results can be tested and exploited in quantitative plant resistance breeding strategies.\\ 

We showed that the quick pathogen adaptation can be explained by the simultaneous restoration of both traits, if the pathogen population has a high reproduction rate, and mutations have a high deviation. Previous studies have shown that the impact of mutations on individual fitness in a specific environment can be described by some particular statistical functions \citep{Mar2006,San2010b}. These functions look like normal laws for various species, e.g. {\it Drosophila melanogaster}, {\it Caerorhabditis elegans}, {\it Saccharomyces cervisiae} or {\it Cryptococcus neoformans} \citep{Mar2006}, but are more complicated for viruses, with functions somewhere between normal and uniform laws \citep{San2010b}. Our model shows that normal mutation law results in slightly faster adaptation to an unfavorable environment than the truncated uniform law, since the distribution tails of a normal law allow, albeit with a low probability, mutations with a strong effect to occur. Nevertheless, these two laws lead to adaptation curves with the same shape, which can be explained by the symmetry of these laws and their constant parameters. The uniform law considered with scaling boundaries leads to progressive deceleration of the adaptation process as shown in \citet{Nov1995}. 

\subsection{The robustness of the proposed stochastic approach for modeling the progressive adaptation of the pathogen population}
The birth and death processes are powerful tools for modeling the adaptation dynamics of pathogens, since they can easily be adapted to many biological situations by adjusting the transition rates \citep{All2003,Nov2006}. Their extension to an infinite state space makes it possible to model life history-trait evolution \citep{Cha2006}. This approach allows us to monitor the stochastic dynamics of mutants, and then to follow the evolution of pathogen traits, even if the genetic mechanisms underlying the adaptation process are poorly understood. Note that the results of the model can be affected by transition functions. However, a comparative analysis of the impact of various transition functions on the dynamics of the model falls outside the scope of our study. To preserve the simplicity of the model, we did not explicitly consider genetic recombination between different pathogen strains. Nevertheless, since our mutations were modeled by changes in the evolution coefficient, they can also be considered to be an implicit mimic of recombination processes.\\

\citet{Rou2008} showed that the evolution coefficient rises logarithmically with the number of lesions until it becomes extremely large. This can be explained by the fact that a beneficial mutation may not go on to be fixed, because of interference with another mutation with a greater beneficial effect that arises either shortly before or shortly after the first mutation \citep{Bar1995}. Our model confirms that if the carrying capacity of the pathogen increases, the adaptation speed does not increase proportionally, but logarithmically. Moreover, the model confirms findings reporting that a mutation with a stronger effect, i.e. which corresponds to the extension of the boundaries of the mutation law in our model, produces an overall reduction in the adaptation time \citep{Met2000}. The model shows that if a pathogen adapts, then the strain with the highest carrying capacity will eventually be selected despite the fact that lesions with the highest reproduction rate have a selective advantage at the beginning of the adaptation process. This finding is consistent with findings of theoretical studies revealing that a quantitative form of resistance and imperfect vaccines select pathogens with higher levels of intrinsic virulence, as measured from the induced host mortality \citep{Gan2000,Gan2001,Mas2006}. It is also consistent with empirical studies showing an increase in the severity of disease induced by pathogens in partially-resistant cultivars in plant epidemiology \citep{Cow2002,Caf2013}. Finally, the model confirms the intuitive assumption that the existence of a trade-off between two traits under selection would decelerate the adaptation process. Analysis of the model showed that an increase in the pathogen reproduction rate or in its carrying capacity will increase the adaptation speed in the five adaptation scenarios considered, but will not change the order of the adaptation curves. Thus, we can conclude that the model's behavior is robust, and that the conclusions drawn about how to manage the progressive adaptation of pathogens are credible.

\subsection{Conclusion}
Using a stochastic model, we studied the durability of quantitative plant resistance from the viewpoint of its effects on the reproduction rate and carrying capacity of the pathogen population. Our theoretical framework can help breeders to develop principles for the sustainable deployment of quantitative trait loci. In particular, the model suggests that in order to slow down progressive pathogen adaptation, QTLs should target pathogen traits whose restoration will be subjected to a trade-off. Existence of putative trade-offs among pathogen traits should be the first indication of their potential antagonistic restoration if they are depressed by QTLs. Moreover, QTLs that decrease the pathogen carrying capacity must be coupled with QTLs that decrease the spore production per lesion, the infection efficiency or those that increase the latent period. Note that our model is appropriate for systemic diseases as well. Our model can also be applied to any situation in which individuals evolve continuously in order to adapt to an environment that has depressed their fitness as a result, for example, of global climate change \citep{Hof2011}.

\section*{Acknowledgment}
Funding for this research was provided by R\'egion Pays de la Loire (MODEMAVE project), and by the SPE Department of INRA.
 
 \vspace*{.5in}
 
\bibliography{biblio}
\newpage
\renewcommand{\baselinestretch}{1.5}
\section*{Figure Legends}
\clearpage
\newpage
\clearpage
\newpage

\begin{table}[h]
\caption{{\small {\bf Definition of variables and parameters used to model pathogen continuous adaptive dynamics.}}}
\begin{tabular}{p{2.5cm}p{2.5cm}p{9cm}}
\hline
\centering Name & \centering Value & Description\\
\hline
\centering $I_n(t)$ & \centering $\mathbb{N}$ & Number of lesions of strain $n$ on the host at time $t$ \\
\centering $I(t)$ & \centering $\mathbb{N}$ & Total number of lesions on the host at time $t$ \\
\centering $N(t)$ & \centering $\mathbb{N}$ & Total number of strains at time $t$ \\
\centering $r$ & \centering $\{0.5,1,2,3\}$ & Pathogen reproduction rate on a susceptible host\\
\centering $K$ &\centering $\{100,500,$ $1000,2000\}$ & Pathogen carrying capacity on a susceptible host\\
\centering $\beta_r^i$ &\centering $]0;1]$ & Evolution coefficient of trait $r$ of pathogen lesion $i$ at time $t$\\
\centering $\beta_K^i$&\centering $]0;1]$ & Evolution coefficient of trait $K$ of pathogen lesion $i$ at time $t$\\
\centering $\vec{\beta^i}= (\beta_r^i,\beta_K^i)$ &\centering $\mathcal{X}= ]0;1]^2$ & Evolution vector of pathogen lesion $i$ at time $t$\\
\centering $\omega$ &\centering $0.05$ & Mutation probability of a pathogen lesion \\
\centering $ \vec{z} = (z_r,z_K)$ & \centering $\mathcal{X}$ & Increment of the evolution coefficient\\ 
\centering $b$ &\centering $\{0.1, 0.2,0.3\}$ & Maximum rise or fall of the evolution coefficient, induced by a mutation\\
\centering $t_y^k$&\centering $\mathbb{R}$ & Time of the $y$th event in the simulation run $k$\\
\centering $\bar{t_y}$&\centering $\mathbb{R}$ & Mean time of the $y$th event over 1000 runs\\
\centering $I^k(t_y^k)$ & \centering $\mathbb{N}$ & Size of the pathogen population at time $t_y^k$ in the simulation run $k$\\
\centering $\bar{I}(\bar{t_y})$ & \centering $\mathbb{N}$ & Mean size of the pathogen population at time $t$ over 1000 runs\\
\centering $\bar{\beta_\tau^k}(t_y^k)$&\centering $]0;1]$&Evolution coefficient of trait $\tau\in\{r,K\}$ of the run $k$ at time $t_y^k$\\
\centering $\bar{\bar{\beta_\tau}}(\bar{t_y})$&\centering $]0;1]$&Mean evolution coefficient of trait $\tau\in\{r,K\}$ at time $\bar{t_y}$ over 1000 runs\\
\hline
\end{tabular}
\label{param}
\end{table}
\end{document}